\documentclass{PoS}
\usepackage{epsfig}
\PoS{PoS(BDMH2004)056} 

\title{Nearby early--type galaxies with ionized gas. \\ Characterization of the underlying stellar population}

\ShortTitle{Nearby early--type galaxies with ionized gas}

\author{\speaker{Francesca Annibali}, Luigi Danese\\
        SISSA, Italy\\
        E-mail: \email{annibali@sissa.it}; \email{danese@sissa.it}}

\author{Roberto Rampazzo, Alessandro Bressan \\
        Osservatorio Astronomico di Padova, Italy\\
        E-mail: \email{rampazzo@pd.astro.it}; \email{bressan@pd.astro.it}}

\author{Emanuele Bertone, Miguel Chavez \\
         INAOE, Puebla, Mexico\\
	 E-mail: \email {ebertone@inaoep.mx}; \email{mchavez@inaoep.mx}}

\author{Werner W. Zeilinger \\
        Institut f\" ur Astronomie der Universit\" at  Wien, Austria\\
        E-mail: \email {zeilinger@astro.univie.ac.at}}

\abstract{We present a preliminary analysis of the sample of 
early type galaxies of Rampazzo et al. (2005), selected to build
a data-set of spectral properties
of well studied  early-type galaxies showing emission lines.
Because of the presence of emission lines,
the sample is biased toward objects that might be expected to have ongoing 
and recent star formation. 
We have compared the line-strength indices presented in Rampazzo et al. (2005)
with Simple Stellar Populations (SSPs) in order 
to characterize the underlying stellar population of the galaxies.
We have derived ages, metallicities and [$\alpha/Fe$] ratios.
The positive trend of the $\sigma$-metallicity and $\sigma-[\alpha/Fe]$ 
relations is reproduced. The bulk of the galaxies span a range in
metallicity from $\sim$ solar to $\sim$ twice solar and a range in
$[\alpha/Fe]$ from $\sim$ 0.2 to $\sim$ 0.4.
Furthermore the comparison of the derived 
parameters at different 
galactocentric distances shows the presence of negative metallicity gradients
from the center outwards.}

\FullConference{Baryons in Dark Matter Halos\\
		 5-9 October 2004\\
		 Novigrad, Croatia}

\begin{document}

\section{Introduction}

\vspace{-0.2cm}

Although  Early-Type galaxies (Es) appear as a homogeneous class, 
characterized by well defined scaling relations like 
the Fundamental Plane, 
the color-magnitude relation and the Mg2-$\sigma$ relation,
much evidence suggests that a secondary episode of star formation 
has occurred during 
their evolutionary history. Simulations indicate that galaxy collisions, 
accretion and merging episodes are important factors
in the evolution of galaxy shapes (see e.g. Barnes (1996);
Schweizer (1996)) and can interfere with their passive evolution.
This understanding of early-type galaxy formation has been
enhanced by the study of interstellar
matter. This component and its relevance in  secular galactic
evolution was widely neglected in early studies of early-type
galaxies since they were for a long time considered to be essentially
devoid of interstellar gas. In the last two decades, however, multi-wavelength
observations have changed this picture and have detected the presence of a
multi-phase Inter Stellar Medium (ISM).
For what concerns the ``warm'' ionized component of the ISM, 
the main ionization mechanism, which does not 
seem to be powered by star formation, remains still uncertain. 
With the aim of improving the understanding of the 
nature of the ionized gas in early-type galaxies,
Rampazzo et al. (2005) (hereafter R05)
have started a study of a sample of Es showing emission 
lines in their spectra. The adopted strategy is to investigate 
the physical conditions of the ionized gas, the possible
ionization mechanisms and the connection
to the stellar population of the host galaxy.

\vspace{-0.1cm}

\section{Observed properties of the Sample}

\vspace{-0.2cm}

The R05 sample 
contains  50 early--type galaxies. The sample is selected
from a compilation of nearby galaxies 
($z<5500 \ km~s^{-1}$) showing ISM traces in at least one of the
following bands: IRAS 100 $\mu$m, X-ray, radio, HI and CO, and  
should be
biased towards objects that might be expected to have ongoing and recent
star formation because of the presence of emission
lines. The emission should comes from a combination of active galactic 
nuclei and star formation regions within the galaxies.
The spectra, taken along the major axis, are provided at different 
concentric regions (apertures) at radii of 1.5",
2.5", 10", r$_e$/10, r$_e$/8, r$_e$/4 and r$_e$/2,  
and in four adjacent regions (gradients) at 
0 $\leq$ r $\leq$r$_e$/16 ("nuclear"), 
r$_{e}$/16 $\leq$ r $\leq$r$_e$/8, r$_{e}$/8 $\leq$ r $\leq$r$_e$/4 and
r$_{e}$/4 $\leq$ r $\leq$r$_e$/2.
For all the apertures and gradients R05  measured  the 
21 line-strength indices of the original Lick-IDS system using the  redefined
passbands of Trager et al. (1998) (TR98) plus the higher order Balmer lines 
introduced by Worthey \& Ottaviani (1997) (WO97). 
The measured indices have been
conformed to the Lick-IDS System following the 
procedure described in WO97.
The correction procedure include correction for possible hydrogen emission, 
correction for velocity dispersion and 
transformation to the Lick system.
For details on the observations and reduction procedure
we refer to R05.
%\section{H $\beta$ emission}

\begin{figure}

\vspace{-0.8cm}
\epsfig{figure=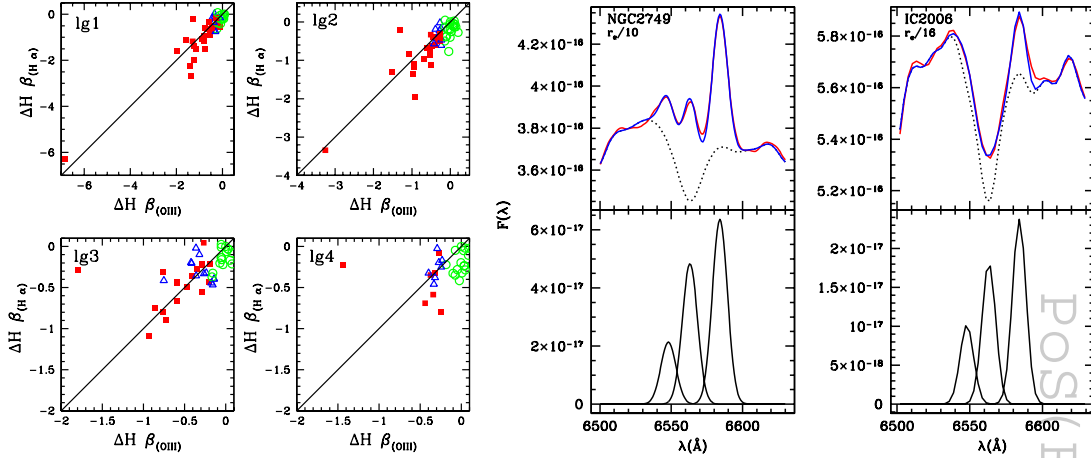, width=1.\textwidth}

\vspace{-0.5cm}

\caption{(left four panels) R05 sample:
 $H\beta$ emission estimate derived from H$\alpha$ against the emission 
derived from $[OIII]$ for the four gradients. 
The solid line is the one-to-one relation.
Open green circles are for galaxies with 
O[III] emission detected under 1$\sigma$ level, blue triangles and red full squares between 1 and 2$\sigma$ levels and above 2$\sigma$ level respectively. 
(right four panels) Fitting of N[II]($\lambda$ 6548, 6584)
and H$\alpha$ lines for two representative galaxies: 
one with H$\alpha$ in emission 
(NGC~2749) and the second with the H$\alpha$ infilling (IC~2006). 
Lower panels show the residuals lines after the subtraction 
of the H$\alpha$ line of the template galaxy NGC~1426 
(dotted lines in the upper panels).}
\label{fig1}
\end{figure}

The presence of emission lines affects the measure of some line--strength
indices. In particular, the H$\beta$ index measure of the underlying stellar
population could be contaminated by a significant infilling due to presence of
the H$\beta$ emission component. 
R05 have measured H$\beta$ emissions in their sample following 
two different approaches:
1) measuring the [OIII] emission and deriving the H$\beta$ emission
 through the relation $EW(H\beta_{em})/EW([OIII]\lambda 5007) = 0.7$ 
found by Gonzalez et al. (1993) for his
sample of early-type galaxies;
2) measuring the H$\alpha$ emitted flux and 
deriving the  H$\beta$ emitted flux according to the relation  $F_{H\beta} =  1/2.86 F_{H\alpha}$ (see e.g. Osterbrock (1989)).
%In both cases, the  emissions have been derived adopting 
%the galaxy NGC~1426 as template for the underlying stellar population. 
In Figure~\ref{fig1} it is shown the comparison  between the two different 
 H$\beta$ emission estimates computed in the four gradients.

%----------------------------Figure1

\vspace{-0.2cm}

\section{Simple Stellar Population Indices}

\vspace{-0.2cm}

Following the procedure described in Bressan et al. (1996),
to which we refer for details, we
have derived line strength indices for SSPs.
In brief all indices are constructed by means of a central
band-pass and two pseudo-continuum band-passes on either side
of the central band (see Worthey et al. (1994) (W94), TR98).
The continuum flux is interpolated between the mid points of
the pseudo-continuum band passes.
The definition of the index in EW is 

\vspace{-0.5cm}

\begin{equation}
\vspace{-0.1cm}
\label{eq1}
I_{EW}= \left(1- F_R/F_C \right) \Delta \lambda
\end{equation}

where $F_R$ and $F_C$ are the fluxes in the line and in the pseudocontinuum
respectively and $\Delta \lambda$ is the width of the central 
band.
The integrated indices for SSPs are calculated according to the following 
method. Given an isochrone in the HR diagram, we derive for each
elementary bin $\Delta \log \frac{L}{L_{\odot}}$ and $\Delta \log T_{eff}$  
the flux in the continuum $F_C$ from libraries 
of stellar spectra and the index I  
from the fitting functions (FFs) of W94 and WO97.
From $F_C$ and I the flux in the central pass-band  $F_R$ is computed inverting
 (\ref{eq1}). After having integrated $F_C$ and  $F_R$ along the 
 isochrone, the final SSP index $I_{SSP}$ is computed from (\ref{eq1}).

\vspace{-0.1cm}

\subsection{$\alpha-enhancement$}

Since the fitting functions are calibrated on Milky Way stars, 
they reflect, at solar
metallicity, the solar element partitions.
Nevertheless several studies have shown that elliptical galaxies 
are characterized by supersolar $[\alpha/Fe]$. 
Models accounting for $\alpha$-enhanced chemical
composition have been presented by Thomas et al. (2003, 2004)  
and Tantalo et al. (2004).
The methods used are based on Response Functions (RFs),
introduced for the first time by  
Tripicco \& Bell (1995),  that describe the behavior
of indices with elements variations.
Similarly to Tantalo et al. (2004), 
we have derived RFs from model atmospheres computed with
the code ATLAS9 for different Z/Z$_{\odot}$ (1/50, 1/5, 3/5), 
T$_{eff}$ (4250, 4500, 5000, 5750, 6250, 7000 K), 
$\log g$ (2, 4, 4.5, 5) and $[\alpha/Fe]$ (0.0 and 0.4), 
in order to construct SSP models with non-solar element partitions.
For fixed T$_{eff}$, g and Z, 
and varying $[\alpha/Fe]$ from 0 to 0.4, the fractional variation 
of the i-th index is: 

\vspace{-0.3cm}

\begin{equation}
R_{i,0.4}=\frac{I_{(i,[\alpha/Fe]=0.4)} - I_{(i,[\alpha/Fe]=0)}} {I_{(i,[\alpha/Fe]=0)}} 
\label{eq2}
\end{equation}

where $I_{(i,[\alpha/Fe]=0.4)}$ and $I_{(i,[\alpha/Fe]=0)}$ are the indices
computed on the synthetic stellar spectra for solar-scaled and 
$\alpha$-enhanced compositions respectively.
At metallicities higher than  $Z/Z_{\odot}=3/5$, 
the responses have been derived through
linear extrapolation.
At this point we computed $\alpha$-enhanced SSPs 
with $[\alpha/Fe]=0.4$ in the following way.
For each elementary bin  
($\Delta \log \frac{L}{L_{\odot}}$, $\Delta \log T_{eff}$) along the isochrone,
we applied a correction $\Delta I$ to the index I derived from the FFs.
The i-th index corrected for the effect of $\alpha$-enhancement is: 

\vspace{-0.3cm}

\begin{equation}
I_{i,0.4} = I_{i,0} + \Delta I_{i,0.4} = I_{i,0} + R_{(i,0.4,T,g,Z)} \times I_{i,0} 
\end{equation}

where $I_{i,0}$ is the value from the FFs and $R_{i,0.4,T,g,Z}$ is obtained 
by linearly interpolating the  $R_{i,0.4}$ values of (\ref{eq2}) at  
the gravity, temperature and metallicity of the bin. 
Once computed the corrected index I, the procedure 
to obtain the final SSP index is the same as  
described at the beginning of the section for solar-scaled models.
More details can be found in Annibali et al. (2005) in prep.

\vspace{-0.4cm}

\section{Results}

\vspace{-0.2cm}

\begin{figure}
\vspace{-0.7cm}
\centering
\epsfig{figure=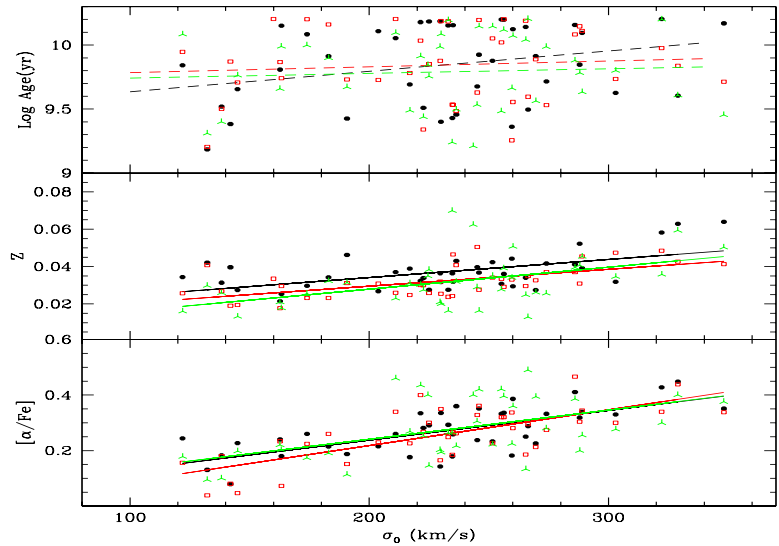, width=11cm,height=7.3cm}

\vspace{-0.3cm}

\caption{From top to bottom we plot the derived ages, Z and $[\alpha/Fe]$
as function of central velocity dispersion $\sigma_0$. 
Black dots, red squares and green triangles 
correspond to
the regions 0 $\leq$ r $\leq$r$_e$/16, r$_{e}$/16 $\leq$ r $\leq$r$_e$/8
 and r$_{e}$/8 $\leq$ r $\leq$r$_e$/4.
Same colors are used for lines which indicate the fits 
to the data.}
\label{fig2}
\end{figure}

We have analyzed a sample of early-type galaxies showing emission lines 
in their spectra (R05) by means of new
SSP indices that account also for the enhancement of the $\alpha$-elements.
It is known that narrow band indices constitute a powerful tool to 
disentangle age, metallicity and enhancement effects 
(see e.g. Worthey (1994), WO97).  
We have devised a simple
but robust algorithm to derive these parameters from the analysis
of the Mgb, [MgFe] and H$\beta$ (or H$\gamma$) indices 
(Annibali et al. (2005) in prep.).
The results of this preliminary investigation are shown in Figure \ref{fig2}.
From top to bottom we plot the derived ages, metallicities and [$\alpha/Fe$]
ratios as function of central velocity dispersion $\sigma_0$.
Black dots, red squares and green triangles correspond respectively to the 
nuclear region (r $\leq$r$_e$/16), to
r$_{e}$/16 $\leq$ r $\leq$r$_e$/8 and to r$_{e}$/8 $\leq$ r $\leq$r$_e$/4.
Same colors are used for lines which indicate the fits to the data. 
The positive trend of
the $\sigma$--Z and $\sigma$- $[\alpha/Fe]$ relations of 
early-type galaxies is reproduced for our sample.
The  galaxies span a range in metallicity from $\sim$ solar
(Z $\sim$ 0.02 ) to $\sim$ twice solar. 
The  $[\alpha/Fe]$ ratios are supersolar and 
go from $\sim$ 0.2 to $\sim$ 0.4
from the smallest to the largest systems. 
The shift between the metallicity relations for the plotted regions
indicates negative Z gradients from the center outwards.
The broad range of ages derived from this analysis is at variance with the 
quite
tight relations obtained for the global metallicity and $\alpha$-enhancement.
This could suggest that the reliability of the whole process of age 
determination by means of Hydrogen indices 
(possibly affected by emission contamination or 
by $\alpha$- enhancement) is still far from 
being satisfactory. Indeed
there are several data with too high ages.
However the spectra are of high quality and the adopted lines are not heavily 
contaminated by emission,
while effects of enhancement should include a bias instead of dispersion. We 
thus conclude that this is an evidence that the formation process of the 
bulk of the galaxy follows a well defined path (and timescale), and that the 
broad range in age is caused by
subsequent  "rejuvenation" episodes involving only a tiny fraction of the 
galaxy mass (e.g. Bressan et al. (1996)).
Thus the "indirect" signature  left on the chemical enrichment path may be
a more robust estimator of the formation timescales than the "direct" 
estimate based on H lines, marking the secular evolution of galaxies.

%!=======================================================================

\end{document}